\documentclass[twoside,slac_one]{revtex4}
\usepackage{graphicx}
\usepackage{fancyhdr}
\usepackage{amsmath} 
\usepackage{bm}
\usepackage{amsxtra}
\usepackage{amssymb}
\usepackage{amsthm}
\usepackage{latexsym}
\usepackage{lscape}
\usepackage[normalem]{ulem}

\newcommand{\Zmumu}   {\mbox{${\mathrm Z}(\rightarrow\mu \mu$)}}

\newcommand{\Zllj}   {\mbox{${\mathrm Z}(\rightarrow l l$)+jets}}
\newcommand{\Zll}   {\mbox{${\mathrm Z}(\rightarrow l l$)}}
\newcommand{\ttbar}   {\mbox{${\mathrm t}\bar{\mathrm t}$}}
\newcommand{\Znunuj}   {\mbox{${\mathrm Z}(\rightarrow\nu \nu$)+jets}}
\newcommand{\Znunu}   {\mbox{${\mathrm Z}(\rightarrow\nu \nu$)}}

\newcommand{\Wmunu}   {\mbox{${\mathrm W}(\rightarrow \mu\nu$)}}
\newcommand{\pt}{\ensuremath{\mathrm{p_T}}}
\newcommand{\met}{\mbox{$\rlap{\kern0.25em/}E_T$}}
\begin{document}

\title{Search for New Physics with Monojet plus missing transverse energy at CMS}

%

\author{Sarah Alam Malik}
\affiliation{The Rockefeller University, NY, USA}

\begin{abstract}
Results are presented for the search for new physics in the monojet plus missing transverse energy channel using $pp$ collision data at a
centre-of-mass energy of 7 TeV. The data were collected by
the CMS detector at the LHC, and correspond to an integrated 
luminosity of $36\pm4$ pb$^{-1}$. In the absence of an excess of events, limits are placed on parameters in the framework of the ADD model and unparticles.
\end{abstract}

\maketitle

\thispagestyle{fancy}


\section{Introduction}

This Letter describes a search for new physics in the missing transverse energy plus single jet final state with $pp$ collisions at a center-of-mass energy of 7 TeV provided by the Large 
Hadron Collider (LHC) at CERN. The data were collected from April 
through November 2010 by the Compact Muon Solenoid (CMS) experiment, 
and correspond to an integrated luminosity of 36 $\pm$ 4 pb$^{-1}$~\cite{ref:EXOPAS}. 

The signature of a single energetic jet recoiling against large missing transverse energy is predicted by numerous theoretical models, one of the most popular being the Large Extra Dimensions model proposed by Arkani-Hamed, Dimopoulos and Dvali~\cite{ref:ADD}. The ADD model aims to explain why gravity is so many orders of magnitude weaker than the other three interactions and in doing so solve the hierarchy problem between the electroweak scale and the Planck scale. The model introduces a number $\delta$ of extra spatial dimensions, which in the simplest scenario are compactified over a torus or sphere with radius $R$. It postulates that the Standard Model (SM) gauge interactions are localised on a (3+1)-dimensional `brane' whilst gravity can propagate in the entire multi-dimensional space, thus rendering its effects as weak in the (3+1)-dimensions.

If we suppose two test masses of mass $m_{1}$ and $m_{2}$ placed within a distance $r \ll R$, the gravitational potential between them is governed by Gauss's law in (4+$\delta$) dimensions and is~\cite{ref:ADD}
\begin{equation}
V(r) \sim \frac{m_{1}m_{2}}{M_{D}^{\delta+2}}\frac{1}{r^{\delta+1}},\,\, (r \ll R)
\end{equation}
where $M_{D}$ is the fundamental Planck scale in the (4+$\delta$)-dimensional theory.
If the two test masses are separated by a distance larger than the radius $R$, the gravitational flux lines can no longer continue to propagate in the extra dimensions and the potential becomes~\cite{ref:ADD}
\begin{equation}
V(r) \sim \frac{m_{1}m_{2}}{M_{D}^{\delta+2}}\frac{1}{R^{\delta}r},\,\, (r \gg R).
\end{equation}
Comparing this to the $1/r$ dependence of the Newtonian potential, we obtain the following relation between the Planck scale in the (4+$\delta$)-dimensional theory and the effective Planck scale ($M_{Pl}$)
\begin{equation}
M_{Pl}^{2} \sim M_{D}^{\delta+2}R^{\delta}.
\end{equation}
The effective Planck scale is thus related to the fundamental Planck scale by the number and size of the extra dimensions. Experimental constraints have excluded the case of $\delta=1$ but for $\delta\ge 2$, setting the fundamental scale to 1 TeV gives extra dimensional distances below $0.5$ mm which can be probed at colliders. 
At these scales, gravity can become stronger than in ordinary space and light Kaluza Klein gravitons can be directly produced. At the LHC, this can occur via a number of initial states; $qg$, $q\bar{q}$ and $gg$. Graviton production in association with a jet produces a distinct topology of a monojet balanced by missing transverse energy which arises from the disappearance of the graviton into the extra dimensions.

Another theoretical model that predicts the monojet plus missing transverse energy signature is unparticle production~\cite{ref:unparticles}. The model predicts a scale invariant new sector which is coupled to the SM through a connector sector with a high mass scale. An operator with a general non-integer scale dimension d$_{\mathrm{U}}$ in a conformal sector induces a continuous mass spectrum of particles. These `unparticles' are here assumed sufficiently long-lived
that they do not decay in the detector and escape without interacting, thus producing missing transverse energy. 
The effects of unparticles below a mass scale $\Lambda_{\mathrm{U}}$ can be studied by using an effective field theory.
While there have been no direct searches, a recent
interpretation of CDF results suggests lower limits on $\Lambda_{\mathrm{U}}$
between 2.11 and 9.19 TeV for $1.05<$d$_{\mathrm{U}}<1.35$ \cite{ref:INTRO_Delgado,ref:INTRO_Kathrein}. 

\section{Event Selection}
\label{sec:eventselection}
The data used to study events with a monojet and missing transverse energy is collected using a combination of jet and $\rlap{\kern0.25em/}E_T$ triggers with the majority of the data being collected by a \met\ trigger with a threshold of 80 GeV. All trigger paths are found to be fully efficient for an offline \met\ cut of 120 GeV. 

Events are required to have atleast one good primary vertex and events originating from cosmic muons, beam-halo and other beam related backgrounds are rejected. 
Jets are reconstructed offline using the particle-flow algorithm~\cite{ref:particleflow}. This algorithm identifies and reconstructs all types of particles produced in the collision by combining information from the tracking system, the calorimeters and the muon system. The reconstructed particles, which include charged hadrons, neutral hadrons, photons, muons and electrons, are fully calibrated and are clustered into jets using the anti-k$_{\rm t}$ algorithm with a distance parameter R = 0.5\cite{ref:antikt}. The momenta of jets are corrected to achieve a uniform relative response of the calorimeter as a function of $\eta$ and an absolute response as a function of \pt. The correction factors are derived from  Monte Carlo (MC) and an additional residual correction factor is obtained from data. The corrected jet momenta are then required to be greater than 30 GeV.

The \met\ is similarly reconstructed using the particle-flow algorithm and defined as the modulus of the negative vector sum of all the reconstructed particles in the event. 

To remove any artificial signal in the calorimeter, a strategy based on unphysical charge sharing between neighbouring channels in space and depth, as well as time and pulse shape, is applied. Signals in a hadronic calorimeter (HCAL) tower or electromagnetic calorimeter (ECAL) crystal identified to be unphysical are removed from the reconstruction.

Whereas most of the beam halo and cosmic muon events are rejected, some of these events leave energy deposits in both the ECAL and HCAL with no corresponding charged track. These events are rejected by requiring that more than 15$\%$ of the energy of the leading jet is assigned to charged hadrons. To reject unidentified electrons and photons, the energy fractions assigned to neutral hadronic, neutral electromagnetic and charged electromagnetic are required to be less than 80$\%$ of the total energy of the leading jet. 
\begin{figure}
  \begin{center}
  \includegraphics[scale=0.39]{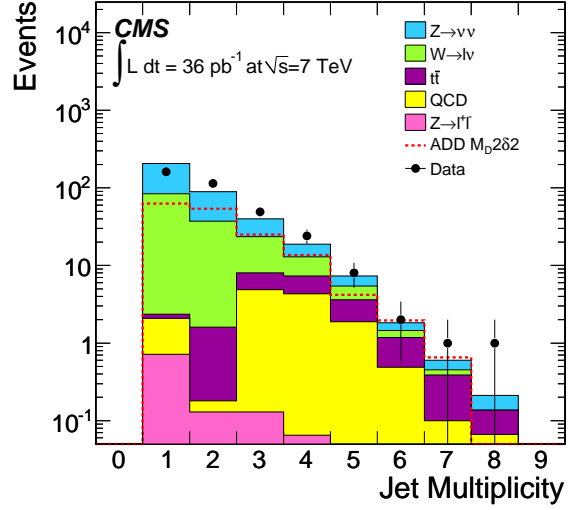}
  \caption{The jet multiplicity distribution for data and the SM backgrounds normalised to the measured rate in data. A representative ADD signal (with M$_D$=2 TeV, $\delta=2$) is shown as a dashed red line.}
\label{fig:njet}
\end{center}
\end{figure}

The search region is chosen by requiring $\met > 150$ GeV and the most energetic jet in the event to have $\pt > 110 $ GeV and $|\eta| < 2.4$. A second jet with $\pt > 30$ GeV is allowed and the event is vetoed if there are any additional jets in the event with $\pt > 30$ GeV. This significantly reduces background from \ttbar\ events and QCD multijets, as is shown in the jet multiplicity distribution in Figure~\ref{fig:njet}, where all other cuts have been applied. In order to suppress background from QCD dijet events, the angular separation between the leading and sub-leading jets is required to be $\Delta\phi(j_{1},j_{2}) <$ 2.0. Events originating from electroweak processes are removed by vetoing the event if one or more well reconstructed and isolated electrons or muons are found. The criteria used to select electrons and muons is described below. In addition to this veto, events with an isolated track with $p_{T} > 10 $ GeV are also rejected. A track is considered isolated if the scalar sum of the transverse momentum of all tracks with $\pt > 1$ GeV in the annulus of $0.02<\Delta R<0.3$ around its direction is less than 10\% of its \pt.

Muon candidates are required to have $\pt > 20$ GeV within $|\eta| < 2.1$ and be reconstructed with compatible track segments in the silicon tracking detectors and the muon detectors. The track formed using hits on these two track segments is required to be of good quality~\cite{ref:muon}.
Electron candidates are reconstructed by matching an energy deposit in the ECAL to hits in the silicon tracker. They are required to have $|\eta| < 1.44$ or $1.56 < |\eta| < 2.5$ to ensure well instrumented regions of the detector. In addition, candidates with a significant mismeasurement in the ECAL or that are consistent with those from photon conversions are rejected~\cite{ref:electron}. 
Both muon and electron candidates are required to originate within 2 mm of the beam axis in the transverse plane and be spatially separated from jets by at least $\Delta R=\sqrt{(\Delta \eta)^2+(\Delta \phi)^2}=0.5$, where $\Delta\eta$ and $\Delta\phi$ are the separation in pseudorapidity and azimuthal angle (in radians) between the muon (electron) and the jet directions respectively.
The muon and electron candidates are also required to be isolated using a variable representing the combined relative isolation, $RelIso$. This is defined as the scalar sum of the transverse momenta of tracks and transverse energies in the ECAL and HCAL in a cone of radius $R=0.3$ around the electron (muon) track, excluding the energy contribution from the candidate, divided by its \pt. Muon candidates are required to have $RelIso < 0.15$ and electrons in the central (forward) regions of the detector are considered isolated if $RelIso < 0.09(0.04)$.

\begin{figure}
  \begin{center}
  \includegraphics[scale=0.39]{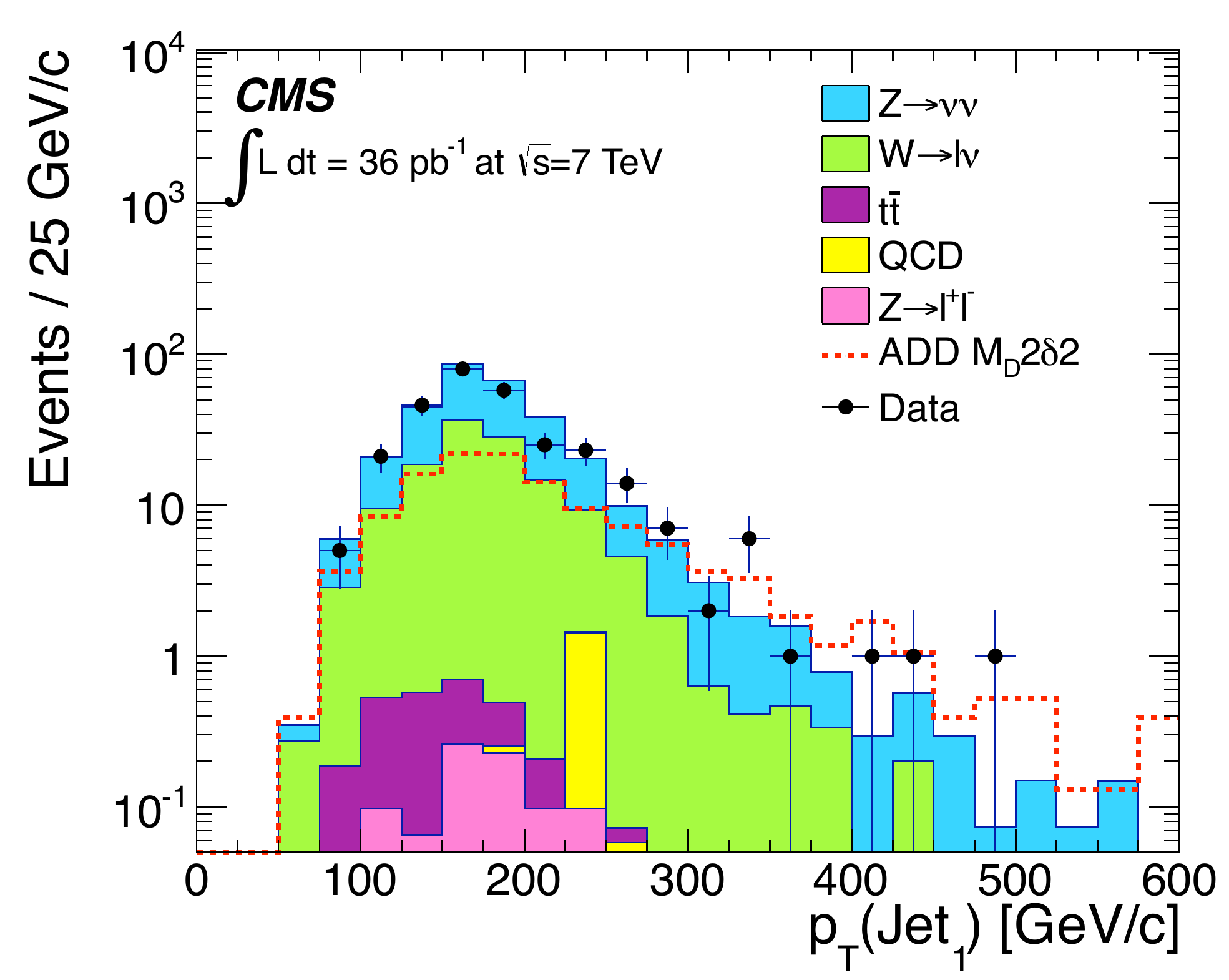}
  \includegraphics[scale=0.39]{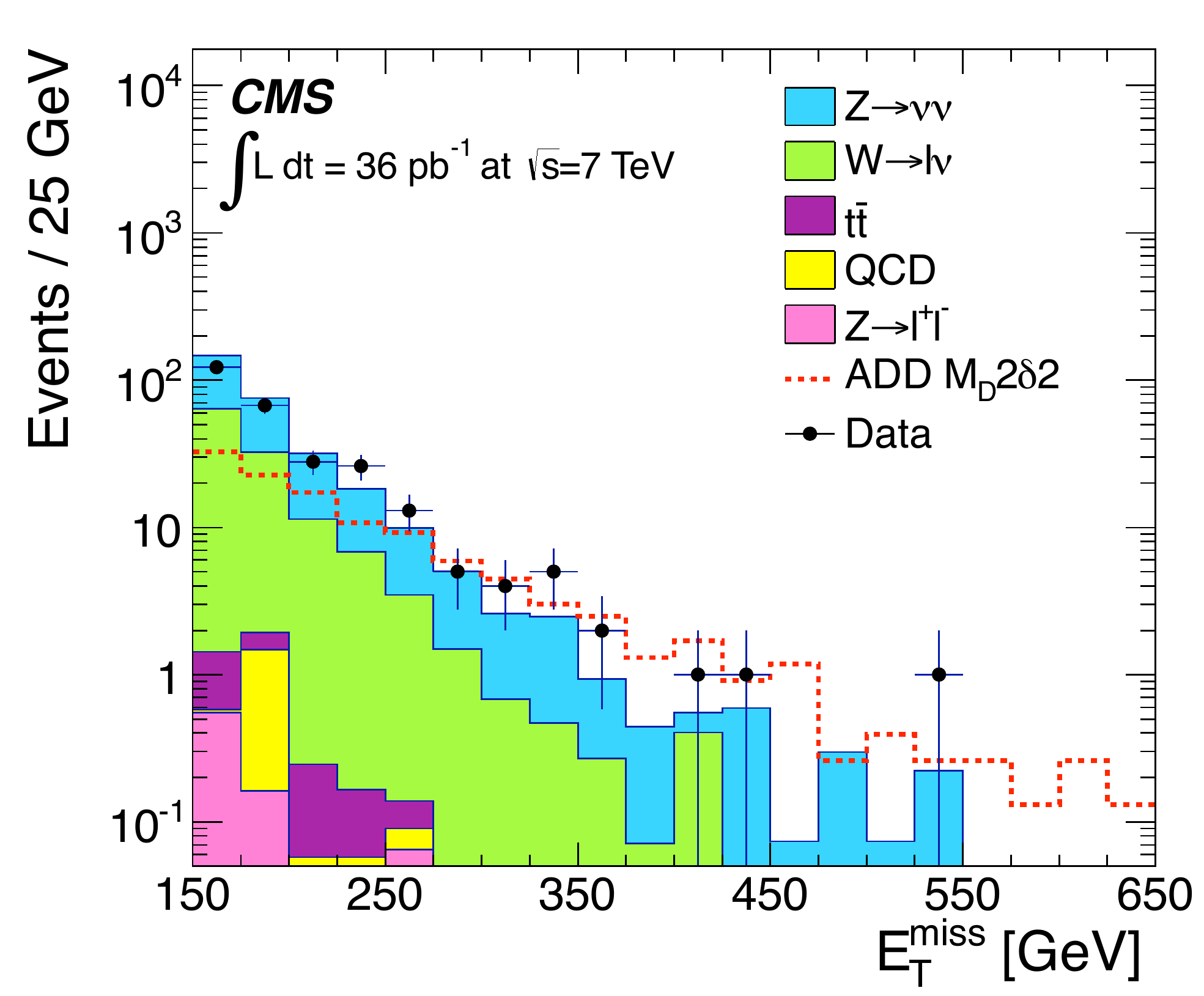}
  \caption{The transverse momentum distribution of the leading jet (left) and the missing transverse energy \met\ distribution (right) are shown for data and the SM backgrounds normalised to the measured rate in data. A representative ADD signal (with M$_D$=2 TeV, $\delta=2$) is shown as a dashed red line.}
\label{fig:jetmetplot}
\end{center}
\end{figure}

The number of events obtained at each stage of the selection process is shown in Table~\ref{tab:yields_dataMC}, where the Standard Model background estimates are taken from Monte Carlo.
\begin{table}[ht]
\begin{center}
\begin{tabular}{lccccc|cc} \hline
Cut & $W$+jets& \Znunu+jets& \Zll+jets & \ttbar & QCD & Total MC & Data\\\hline
$\met\!>\!150$ GeV, jet cleaning
                     & 622   &  259  & 46.7   &  90.4   & 202     & 1220 & 1298\\
$\pt\!(j_1)>\!110$ GeV , $|\eta(j_1)|\!<\!2.4$
                     & 583   &  245  & 43.4   &  76.9   & 201     & 1149 & 1193\\
N$_\textrm{jets}\le 2$
                     & 446   &  201  & 34.3   &  11.3   & 74.3      & 767  & 778 \\
$\Delta\phi(j_1,j_2)<2$
                     & 370   &  182  & 29.5   &  9.1    & 6.3       & 597  & 596 \\
Lepton Removal
                     & 107   &  173  & 0.8    &  1.7    & 1.4       & 284  & 275 \\\hline
\end{tabular}
\caption{Event yields at each stage of the selection process, using leading order MC normalised to the integrated luminosity of the data.}
\label{tab:yields_dataMC}
\end{center}
\end{table}
Table~\ref{tab:yields_dataMC} shows that after the full event selection, the only significant remaining backgrounds are from electroweak processes where a neutrino(s) in the final state produces real missing transverse energy.
The missing transverse energy distribution and the transverse momentum distribution of the leading jet are shown in Figure~\ref{fig:jetmetplot} for the data, SM backgrounds and a representative ADD signal (with M$_D$=2 TeV, $\delta=2$).
\section{Data-driven background estimation}
The dominant remaining backgrounds from \Znunuj\ and W+jets are obtained using a data-driven method. \\
The associated production of a Z boson with jet(s) and its subsequent decay to a pair of neutrinos constitutes an irreducible background to this search. 
The \Znunu\ background can be estimated by a number of different methods. In this analysis, the \Wmunu\ plus jets sample was used to estimate the background, whilst the statistically limited \Zmumu\ sample provided a useful cross-check.
The \Znunu\ yield is estimated by correcting the \Wmunu\ and \Zmumu\ events for the acceptance, reconstruction efficiencies and the difference in branching fractions and interpreting the muon(s) as missing energy to emulate the missing energy distribution in \Znunu\ events. 

Both the \Znunu\ and the W+jets background estimation utilise a control sample of $\mu$+jets which is obtained from the same data sample as the signal. The search selection is then applied and one or more well reconstructed and isolated muons are explicitly required using the selection criteria described in Section~\ref{sec:eventselection}. 

\begin{figure}
  \begin{center}
  \includegraphics[scale=0.39]{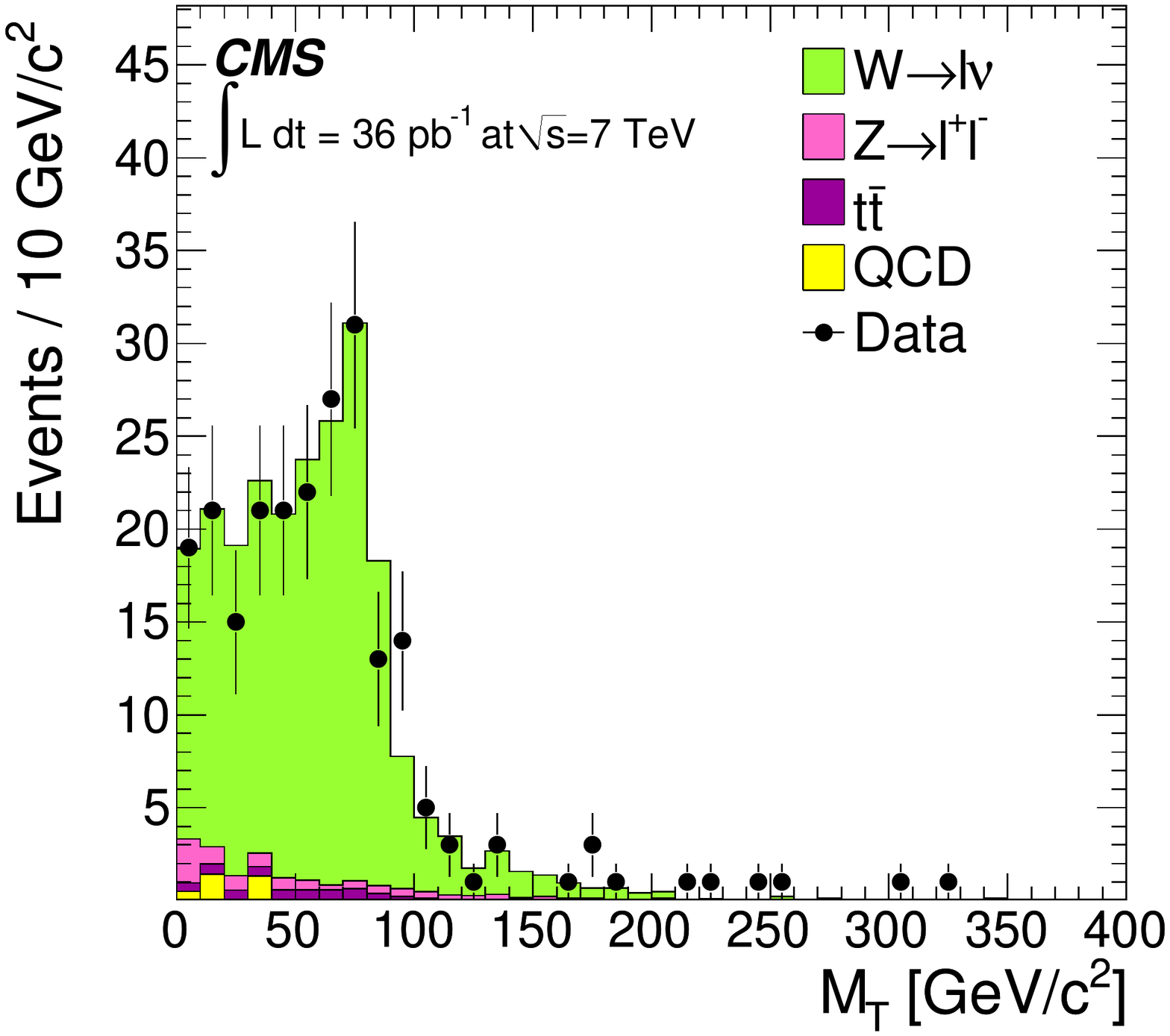}
  \includegraphics[scale=0.39]{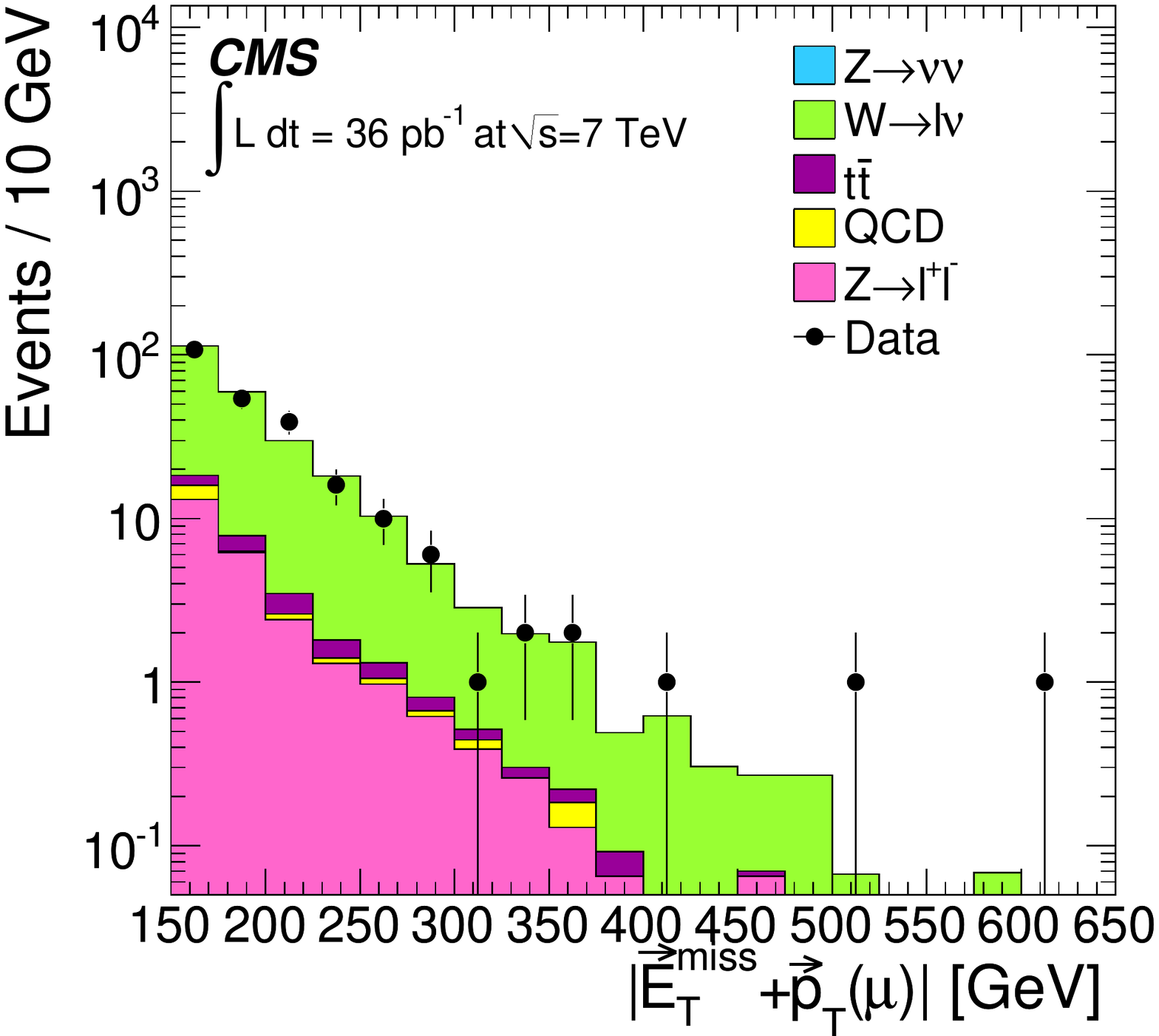}
  \caption{The transverse mass of the $W$ (left) and the missing transverse energy \met\ distribution (right) are shown for data and the SM backgrounds normalised to the measured rate in data. A representative ADD signal (with M$_D$=2 TeV, $\delta=2$) is shown as a dashed red line.}
\label{fig:wmunu}
\end{center}
\end{figure}

In order to select a clean sample of \Wmunu\ events, one well reconstructed and isolated muon is selected and the transverse mass ($M_{T}$), defined as $M_{\rm T}=\sqrt{2\pt^{\mu}\met\left(1-\cos(\Delta\phi\right))}$ where $\pt^{\mu}$ is the transverse momentum of the muon and $\Delta\phi$ is the angle between the the muon and the \met\ vector directions in the transverse plane, is required to be within 50 and 100 GeV. The number of single muon events passing this selection is found to be 113 in the data, compared to 103 expected from MC (95.3 W+jets, 2.9 W($\tau\nu+$jets), 2.4 Z+jets, 2.4 \ttbar\ and 0.08 from QCD multijets). The transverse mass distribution of this sample is shown in Figure~\ref{fig:wmunu} for the data compared with the expectation from MC. The distribution of the vector sum of the missing transverse energy and the muon which emulates the missing energy in \Znunu\ decay is also shown in Figure~\ref{fig:wmunu}. The data is found to agree well with MC expectations.
The number of \Znunu\ events in the signal sample is estimated by correcting the yield by several factors; the reciprocal of the geometric and kinematic acceptance taken from MC and found to be $2.40\pm 0.12$, the lepton selection efficiency in the signal region taken from MC and found to be $0.95\pm 0.02$, the correction for background contributions from non-\Wmunu\ events is found to be $0.923 \pm 0.071$, the ratio of cross-section times branching ratio $\sigma(Z(\mu\mu)+j)/\sigma(W(\mu\nu)+j)\times BR(Z\to \nu\nu)/BR(Z\to \mu\mu)$ = $1/10.76$~\cite{ref:BKG_PDG}$\times 5.95$~\cite{ref:BKG_PDG} =($0.553\pm 0.021$) and an additional factor of $1.33\pm0.03$ is applied to account for the difference in the $p_{T}$ spectrum of the $W$ and $Z$ bosons for $\pt(W,Z)>150$ GeV. All the uncertainties quoted for the above numbers include both statistical and systematic effects.
The number of \Znunu\ events in the signal region predicted from \Wmunu\ events is $176\pm30$.

A sample of \Zmumu\ events are selected by requiring two well reconstructed and isolated muons with opposite sign charge and invariant mass consistent with that of the Z boson. The missing transverse energy of the sample is redefined as the vector sum of the two muons in the event to represent the missing energy in \Znunu\ events. After applying the search selection, 13 \Zmumu\ candidate events are observed with negligible contribution from background. This yield is corrected for the detector acceptance, the muon reconstruction efficiency and the difference in branching ratio between \Znunu\ and \Zmumu\ to obtain a background prediction for \Znunu\ of 162$\pm$45 events, consistent with that from the \Wmunu\ control sample.

The other main background to the search region from W+jets is also estimated using the single muon data sample. It is obtained by scaling the W+jets MC events passing the search selection by the ratio of the observed and predicted \Wmunu\ events in the muon sample. The remaining W+jets background is estimated to be 117$\pm$16, where the uncertainty includes the statistical uncertainty of the muon data sample, the statistics of the MC sample, the uncertainty on the non-\Wmunu\ contribution and the uncertainty on the geometric and kinematic acceptance.

The contribution to the signal region from other backgrounds like QCD, \ttbar\ and \Zllj\ is small and is estimated using MC with an uncertainty of 100\% assigned to the value. 
The expected number of events from all background sources is found to be $297\pm 45$, where the uncertainty reflects the combined statistical and systematic uncertainty with correlations taken into account.

\section{Results}

The absence of an excess of events in the data can be interpreted in the context of the ADD model and unparticle production. The upper limit on the number of non-SM events consistent with the measurement is determined using a Bayesian method with a flat prior for signal and a log-normal density function for the background.

The sources of the systematic uncertainties on the modeling of the ADD signal are;
\begin{itemize}
\item the jet energy scale, estimated by scaling the jet four-vectors by an $\eta$- and \pt-dependent factor~\cite{ref:ANA_JME-10-010}. This yields a variation of 3-7$\%$ for the different ADD scenarios.
\item the jet energy resolution, estimated from a $\gamma$+jet sample~\cite{ref:SYS_JME-10-014}. This results in an uncertainty variation of 0.3-2.2$\%$.
\item PDF uncertainty, evaluated using a reweighting technique. This is found to be 1-2$\%$.
\item a 4$\%$ uncertainty on the luminosity measurement.
\end{itemize}

\begin{figure}
  \begin{center}
  \includegraphics[scale=0.39]{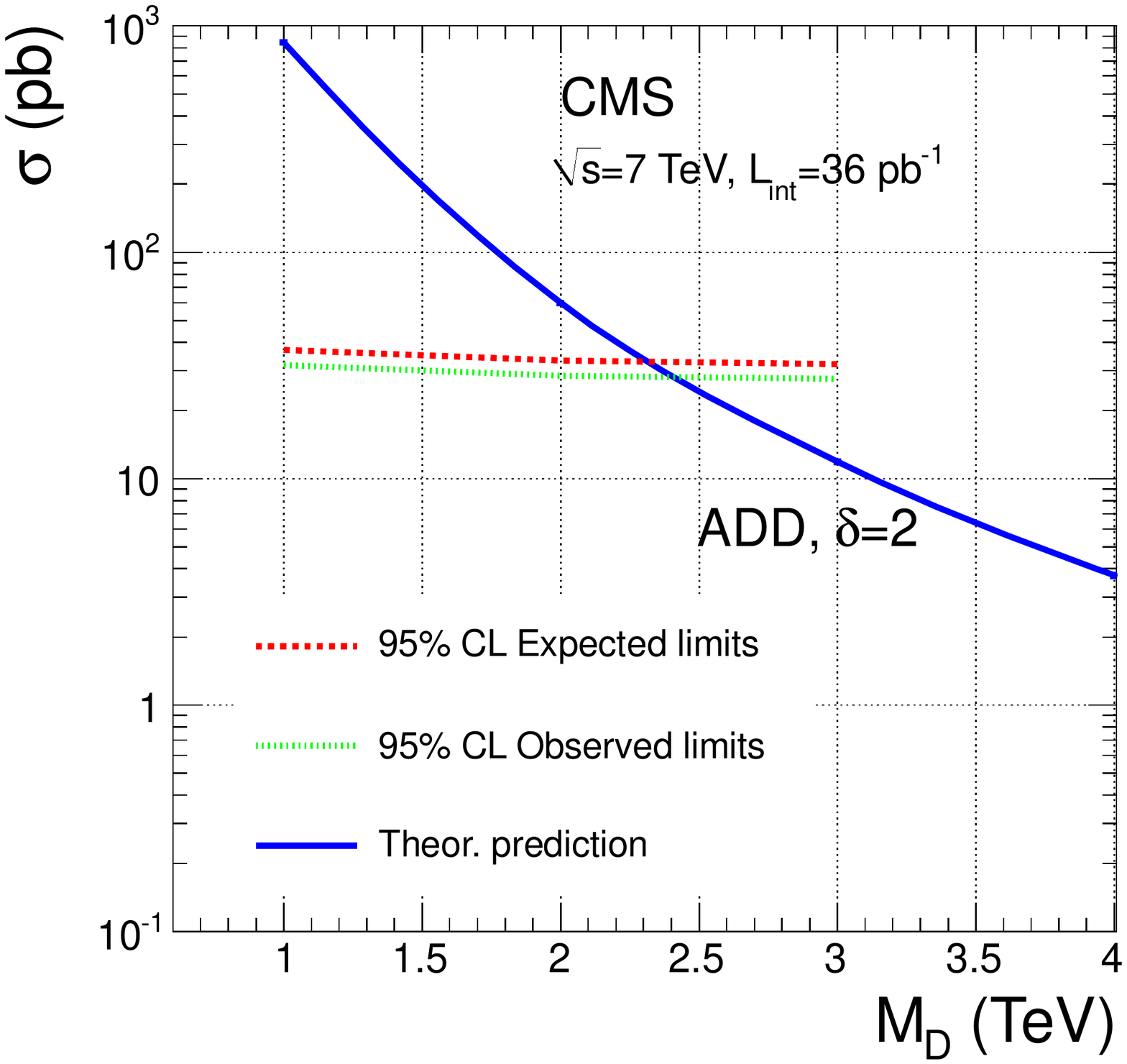}
  \includegraphics[scale=0.39]{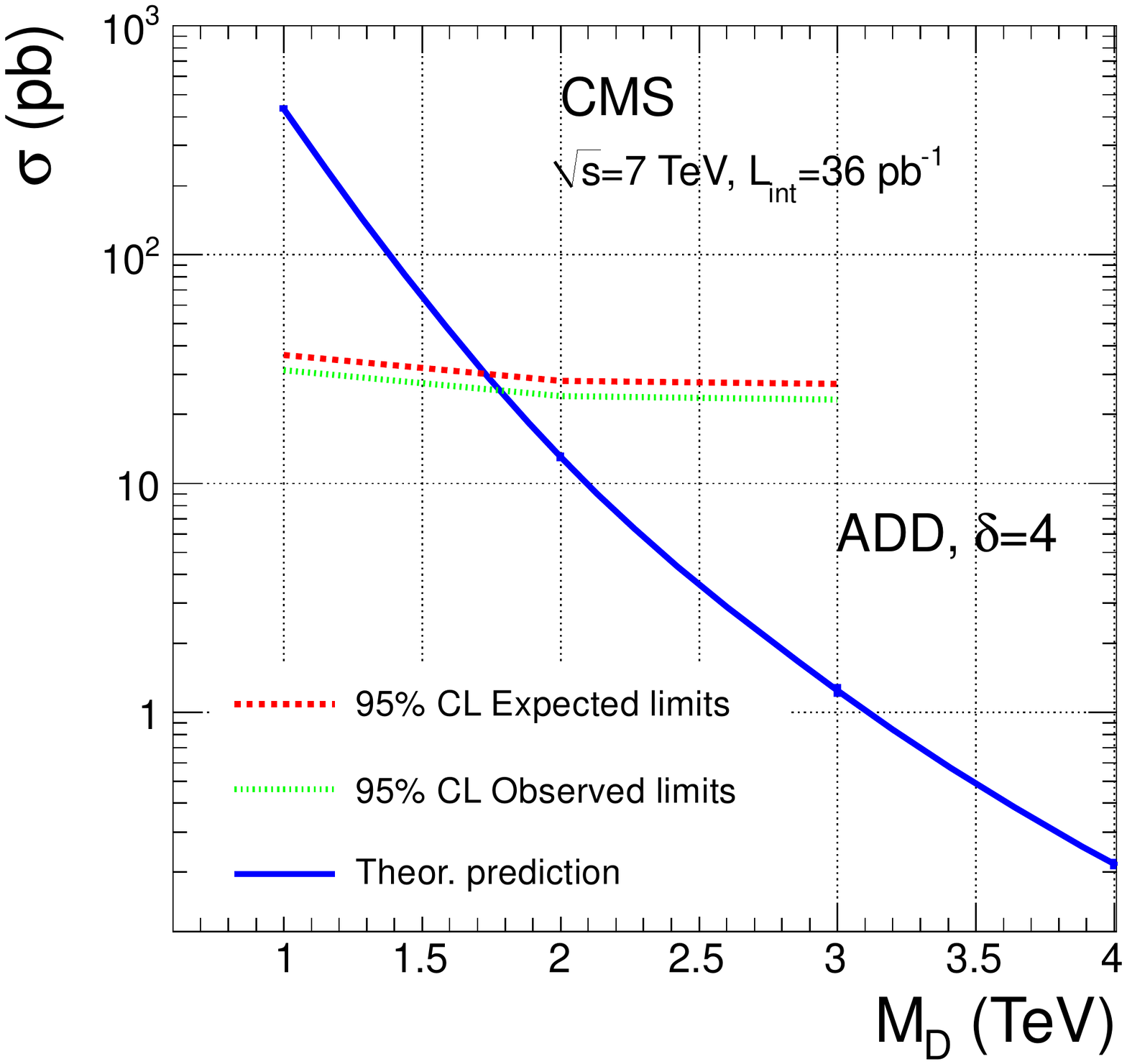}
  \caption{Observed and expected 95\% CL limits on the theoretical cross-section as a function of M$_D$ for the ADD model. The limits are shown for $\delta=2$ and $\delta=4$.}
\label{fig:ADDlimits}
\end{center}
\end{figure}

The exclusion limits at 95$\%$ confidence level (CL) on the fundamental scale $M_D$ in the ADD model as a function of the number of extra dimensions are given in Table~\ref{tab:limits} and shown in Figure~\ref{fig:ADDlimits}.
\begin{table}
        \begin{center}
        \begin{tabular}{c|c|cc|cc} \hline
~$\delta$~ & $K$ factor& LO Exp. & LO Obs. & NLO Exp. & NLO Obs. \\ \hline
 2  & 1.5 & 2.17 & 2.29 & 2.41 & 2.56  \\
 3  & 1.5 & 1.82 & 1.92 & 1.99 & 2.07  \\
 4  & 1.4 & 1.67 & 1.74 & 1.78 & 1.86  \\
 5  & 1.4 & 1.59 & 1.65 & 1.68 & 1.74  \\
 6  & 1.4 & 1.54 & 1.59 & 1.62 & 1.68  \\\hline
                \end{tabular}
        \caption{
Observed and expected 95\% CL lower limits on the ADD model parameter M$_D$
(in TeV) as a function of $\delta$, with and without NLO
$K$ factors applied.
\label{tab:limits}}
\end{center}
\end{table}

\begin{figure}
  \begin{center}
  \includegraphics[scale=0.39]{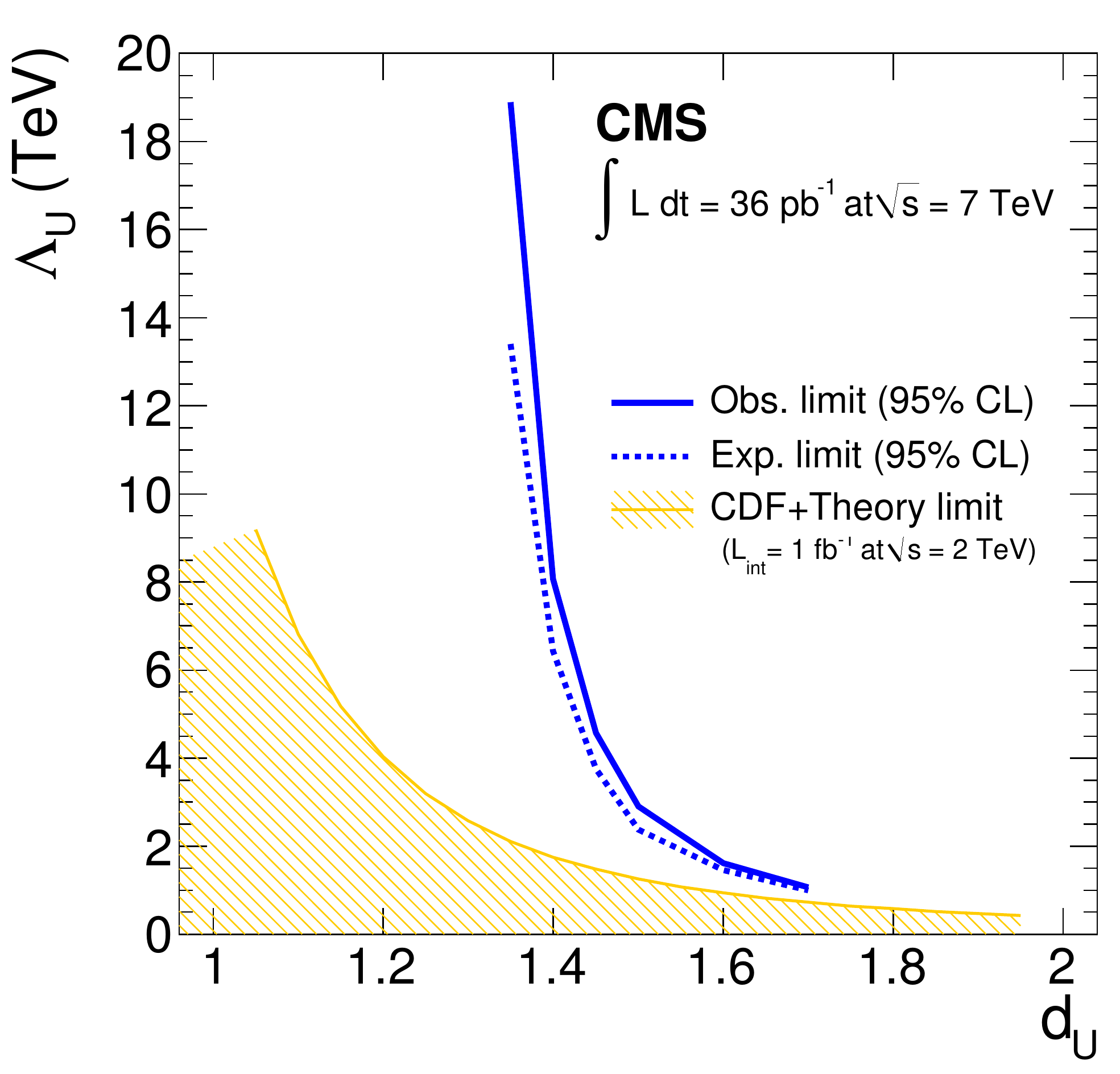}
  \caption{Observed and expected 95\% CL lower limits on the allowed region of unparticle model parameters d$_U$ and $\Lambda_U$, compared to those derived from CDF results~\cite{ref:INTRO_Delgado,ref:INTRO_Kathrein}.}
\label{fig:unparticle_limit}
\end{center}
\end{figure}

The results are also interpreted in the context of the unparticle model and lower limits can be placed at the 95$\%$ confidence level on the unparticle model parameters d$_{\mathrm U}$ and $\Lambda_{\mathrm U}$. The observed and expected 95$\%$ CL lower limits and those derived from CDF results are shown in Figure~\ref{fig:unparticle_limit}. 
The source of systematic uncertainties related to the modeling of the unparticle signal is the same as that for the ADD model with contributions from; the jet energy scale of magnitude 7.5\%--11.5\% for the different unparticle scenarios, the jet energy resolution uncertainty of 0.6\%--2.9\%, PDF uncertainty of 3\%--7\% and a luminosity uncertainty of 4\%.

\section{Summary}

In summary, a search is performed using 36 pb$^{-1}$ of data collected by the CMS experiment in $pp$ collisions at 7 TeV in the monojet plus missing transverse energy channel. The dominant backgrounds after the event selection are from electroweak processes where the Z decays to a pair of neutrinos and the W decays leptonically. These backgrounds are estimated using a data-driven method with a control sample of $\mu$+jets. Other backgrounds from QCD, top pair production and remaining electroweak processes are found to be small and obtained from MC. The data is found to be in agreement with the MC and limits are placed at the 95$\%$ confidence level on the ADD model and unparticles.
 \begin{acknowledgements}
The author would like to thank the accelerator division at CERN for the excellent performance of the LHC machine and providing the data used in this measurement, the CMS collaboration and the organisers of the DPF conference. 
\end{acknowledgements}

\bigskip 

\end{document}